\documentclass{iau} 
\usepackage{graphicx}

\newcommand\aj{AJ} 
\newcommand\apj{ApJ} 
\newcommand\apjs{ApJS}       
 
\newcommand\aap{A\&A} 
\newcommand\mnras{MNRAS} 
\newcommand\apjl{ApJ}

\newcommand\aapr{A\&AR} 
\newcommand\araa{AA\&A}

\title[Multiple stellar populations in M\,54] 
{Multiple populations in the Sagittarius nuclear cluster M\,54 and in other anomalous globular clusters}

\author[A.\,P.\,Milone]   
{A.\,P.\,Milone$^1$
}

\affiliation{$^1$ Research School of Astronomy \& Astrophysics, Australian National University, Mt Stromlo Observatory, via Cotter Rd, Weston, ACT 2611, Australia. email: {\tt milone@mso.anu.edu.au}}

\pubyear{2015}
\volume{317}  
\jname{The general assembly of galaxy halos. Structure, origin and evolution}
\editors{M.\,Arnaboldi, A.\,Bragaglia, M.\,Rejkuba, D.\,Romano eds.}
\begin{document}

\maketitle

\begin{abstract}
M\,54 is the central cluster of the Sagittarius dwarf galaxy. This stellar system is now in process of being disrupted by the tidal interaction with the Milky Way and represents one of the building blocks of the Galactic Halo.\\
 Recent discoveries, based on the synergy of photometry and spectroscopy have revealed that the color-magnitude diagram (CMD) of some massive, anomalous, Globular Clusters (GCs) host stellar populations with different content of heavy elements.
In this paper, I use multi-wavelength {\it Hubble Space Telescope} ({\it HST}\,) photometry to detect and characterize multiple stellar populations in M\,54. I provide empirical evidence that this GC shares photometric and spectroscopic similarities with the class of anomalous GCs.
 These findings make it tempting to speculate that, similarly to Sagittarius nuclear cluster M\,54, other anomalous GCs were born in an extra-Galactic environment.
\end{abstract}

\section{Introduction}
Several studies, based on high-precision photometry, have established that the CMD of nearly all the GCs is made of multiple stellar sequences, that have been identified among all the evolutionary stages (e.g.\,Milone et al.\,2012; Piotto et al.\,2015).
The multiple sequences of the majority of GCs correspond to stellar populations with different content of those light elements involved in H-burning reactions like C, N, O, Na, Mg, and Al (e.g.\,Gratton et al.\,2004, 2012; Marino et al.\,2008; Yong et al.\,2008) and different helium abundance (e.g.\,D'Antona et al.\,2002, 2005; Piotto et al.\,2007; Milone et al.\,2014). Noticeable, while star-to-star variations in light elements and helium have been observed in nearly all the clusters, most of them have homogeneous iron content (e.g.\,Carretta et al.\,2009).
In this context, one of the most-intriguing discoveries of the last years is that a small, but still increasing, number of massive GCs host stellar populations with different metallicity (Marino et al.\,2009, 2011a, 2015; Da Costa et al.\,2009;  Carretta et al.\,2010a,b; Yong et al.\,2014; Johnson et al.\,2015).  

In this paper, I will investigate M\,54, which is a very massive GCs ({\it M} $\sim$2 $\times$ 10$^6$ $M_{\rm \odot}$, McLaughlin \& Van der Marel\,2005) located in the nuclear region of the Sagittarius dwarf spheroidal galaxy (Sgr). 
 Due to its present-day position, it has been speculated that M\,54 is the nucleus of the Sgr dwarf and that the stellar system including M\,54 and the Sgr represents the local counterpart of nucleated dwarf ellipticals (e.g.\,Sarajedini \& Layden\,1995). As an alternative, M\,54 formed in an external region of the dwarf and has been then dragged into the bottom of the Sgr potential well because of decay of the orbit due to dynamical friction (e.g.\,Monaco et al.\,2005; Bellazzini et al.\,2008). 
  In any case, it is widely accepted that M\,54 is associated with the Sgr and that it formed and evolved in the environment of a dwarf spheroidal galaxy.

 In the following, I will present multi-wavelength {\it HST} photometry of M\,54 and of the surrounding Sgr dwarf from the {\it UV legacy survey of Galactic GCs} (Piotto et al.\,2015) and identify the multiple stellar populations within the M\,54$+$Sgr stellar system. 
 Moreover, I will use both spectroscopy and photometry to investigate possible connections between M\,54 and the other anomalous GCs exhibiting metallicity variations. 
\section{Mapping multiple stellar populations in GCs}
\label{sec:mappe}
\begin{figure}[b]
 \begin{center}
  \includegraphics[width=4.45in]{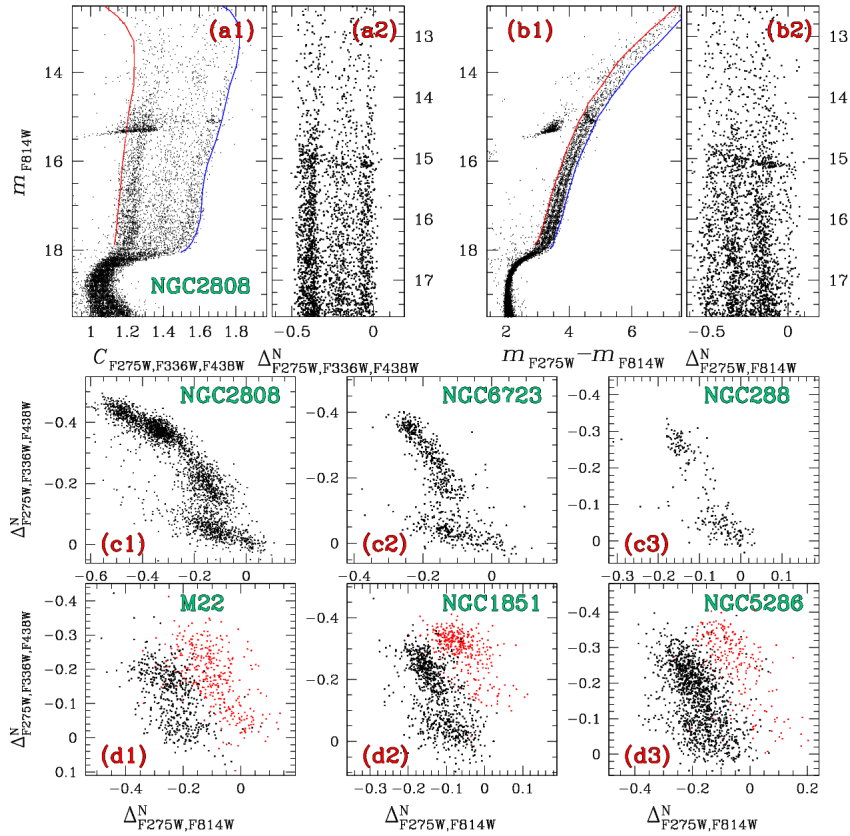}
  \caption{Panels a1 to b2 illustrate the procedure to derive the $\Delta^{\rm N}_{\rm F275W, F336W, F438W}$ vs.\, $\Delta^{\rm N}_{\rm F275W, F814W}$ diagram of RGB stars in NGC\,2808 plotted in the panel c1. 
 Panels c2, c3, d1, d2, and d3 show the same diagram but for the monometallic GCs NGC\,6723 and NGC\,288 and for the anomalous GCs M\,22, NGC\,1851, and NGC\,5286. Stars in the anomalous RGB are colored red  (see text for details). }
    \label{fig:fig1}
 \end{center}
 \end{figure}

Recent papers have revealed that any CMD or two-color diagram made with appropriate combination of filters is very efficient to identify multiple stellar populations in GCs. In particular, the $m_{\rm F275W}-m_{\rm F814W}$ color and the pseudo-color $C_{\rm F275W,F336W,F438W}=$($m_{\rm F275W}-m_{\rm F336W}$)$-$($m_{\rm F336W}-m_{\rm F438W}$) introduced by Milone et al.\,(2013)  are very sensitive to stellar populations with different helium and light-element abundance and have been recently used by Piotto et al.\,(2015) in their survey of GCs to identify stellar populations in a large number of clusters.
As an example, panels a1 and b1 of Fig.~\ref{fig:fig1} show $m_{\rm F814W}$ vs.\,$C_{\rm F275W,F336W,F438W}$ and $m_{\rm F814W}$ vs.\,$m_{\rm F275W}-m_{\rm F814W}$ for NGC\,2808 and clearly reveal its multiple RGB.

 In order to combine information from both diagrams and identify stellar populations within NGC\,2808, Milone et al.\,(2015a) have introduced the method illustrated in the upper panels of Fig.~\ref{fig:fig1}, where the blue and the red fiducial superimposed on each diagram mark the bluest and the reddest envelope of the RGB.
These fiducials are used to verticalize the two CMDs in a way that the blue and the red fiducials translate into vertical lines. The verticalized, $m_{\rm F814W}$  vs.\,$\Delta^{\rm N}_{\rm F275W, F336W, F438W}$ and $m_{\rm F814W}$  vs.\,$\Delta^{\rm N}_{\rm F275W, F814W}$ diagrams are plotted in panels a2 and b2 for RGB stars in NGC\,2808\footnote{ Specifically, Milone et al.\,(2015a) have defined:\\
$\Delta_{\rm X}^{\rm N}= W [(X-X_{\rm blue~fiducial})/(X_{\rm red~fiducial}-X_{\rm blue~fiducial})]-1$,\\  where $X$=($m_{\rm F275W}-m_{\rm F814W}$) or ($C_{\rm F336W,F438W,F814W}$) and $X_{\rm blue~fiducial}$ and $X_{\rm red~fiducial}$ are obtained by subtracting the color of the fiducial at the corresponding F814W magnitude from the color of each star. The constant, $W$, is chosen as the distance between the red and the blue fiducial two F814W magnitudes above the turn off.}  
The resulting $\Delta^{\rm N}_{\rm F275W,F336W,F438W}$ vs.\,$\Delta^{\rm N}_{\rm F275W,F814W}$ plot is shown in panel (c1) for NGC\,2808 and reveals its five distinct populations (see Milone et al.\,2015a for details). Alvio Renzini has named this diagram `chromosome map' and I will keep this nickname in the following.

Panels c2--d3 of Fig.~\ref{fig:fig1} show the chromosome map for five additional GCs. Among them NGC\,6723 and NGC\,288 have homogeneous abundance of iron and neutron-capture elements, similarly to NGC\,2808. On the contrary, M\,22, NGC\,1851, and NGC\,5286 are anomalous GCs and host two distinct groups of stars with different metallicity and s$-$process elements-abundance (Marino et al.\,2009, 2012, 2015; Yong et al.\,2008, 2009, 2015; Villanova et al.\,2010; Lee\,2015; Lim et al.\,2015; Carretta et al.\,2010a).   

 Noticeably, RGB stars in anomalous GCs distribute along two parallel sequences, with the anomalous stars enhanced in heavy elements (red points in Fig.~\ref{fig:fig1}) having, on average, larger $\Delta^{\rm N}_{\rm F275W,F814W}$ values than stars with standard chemical composition. In contrast, the sequence of anomalous stars is not present in monometallic GCs. 
 This finding makes the chromosome map a powerfull tool to identify anomalous GCs from photometry of RGB stars. 

\section{Multiple stellar populations in M\,54}
In the last two decades several papers on the photometry of stellar populations in the system including M\,54 and the Sgr dwarf have been published by different groups (see Siegel et al.\,2007 and references therein). In contrast, very little attention has been dedicated to stellar populations within M\,54. 

 A strong evidence that the CMD of M\,54 is not consistent with a simple population has been provided by Piotto et al.\,(2012) who have discovered that the sub-giant branch (SGB) of this cluster is bimodal in the optical $m_{\rm F606W}$ vs.\,$m_{\rm F606W}-m_{\rm F814W}$ CMD. 
 This is an unusual feature for Galactic GCs, indeed while the SGB of clusters without metallicity variations is narrow and well defined when observed in visual filters, a split or broadened SGB is a distinctive feature of anomalous GCs with variation in [Fe/H] and [(C+N+O)/Fe] (Milone et al.\,2008; Cassisi et al.\,2008; Ventura et al.\,2009; Sbordone et al.\,2011; Marino et al.\,2009, 2012).

 \begin{figure}[b]
 \begin{center}
  \includegraphics[width=4.2in]{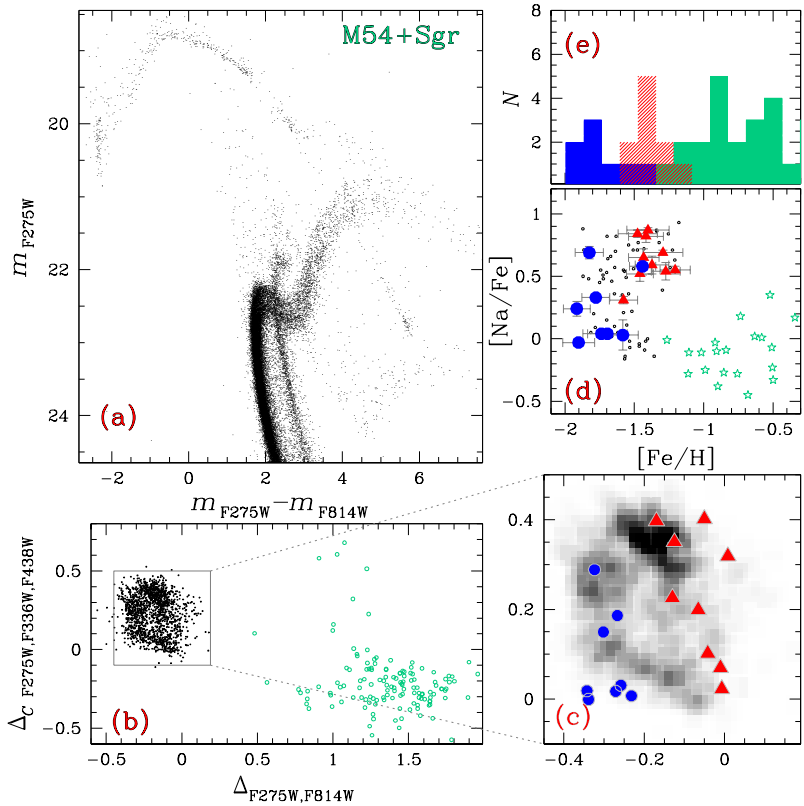}
  \caption{$m_{\rm F275W}$ vs.\,$m_{\rm F275W}-m_{\rm F814W}$ CMD for stars in a $\sim$3$\times$3 region around the center of M\,54 (panel a). Panel b shows the chromosome map for M\,54 RGB stars (black points) and for Sgr stars (aqua circles), while the Hess diagram of M\,54 stars is plotted in panel c. Panel d shows [Na/Fe] vs.\,[Fe/H] from Carretta et al.\,(2010b), and the histograms of the iron distribution is provided in panel e for stars in the standard and anomalous population of M\,54 and for stars in the Sgr dwarf spheroidal galaxy. Blue and red colors in panels c, d, and e indicate normal and anomalous stars in  M\,54, respectively, while Sgr stars are represented with aqua color codes.}
    \label{fig:fig2}
 \end{center}
 \end{figure}
More recently, multi-wavelength photometry from the {\it HST} UV survey of Galactic GCs (Piotto et al.\,2015) has revealed that M\,54 hosts a complex system of multiple stellar populations that are clearly visible along the main sequence, the SGB, the RGB, and horizontal branch in the $m_{\rm F275W}$ vs.\,$m_{\rm F275W}-m_{\rm F814W}$ CMD of Fig.~\ref{fig:fig2}a. The chromosome map of RGB stars plotted in the panel b of Fig.~\ref{fig:fig2} clearly separates the M\,54 members (black points) from the Sgr stars (aqua points). The Hess diagram shown in the panel c for M\,54 reveals that the chromosome map of this cluster consists of two sequences of stars in close analogy with what observed in anomalous GCs (see Fig.~\ref{fig:fig1}). Moreover, the fact that each sequence exhibits two or more distinct clumps demonstrates that both the normal and the anomalous group host sub-populations with different helium and light-element abundance.

Carretta et al.\,(2010b) have derived homogeneous abundances of Fe, O, Na from high-resolution spectroscopy for more than 100 stars in M\,54 and in the surrounding nucleus of the Sgr dwarf galaxy. They have concluded that M\,54 exhibits intrinsic iron dispersion of about 0.2 dex and found a very-extended Na-O anticorrelation among cluster stars.
 Panel (d) of Fig.~\ref{fig:fig2} shows [Na/Fe] vs.\,[Fe/H] from Carretta and collaborators for M\,54 members (black symbols) and for Sgr stars (aqua symbols). 

 For 18 stars both spectroscopy and {\it HST} photometry is available. Among them, the eight stars represented with large blue dots in Fig.~\ref{fig:fig2} belong to the normal RGB, and the ten stars in the anomalous RGB are plotted with red triangles. Stars in the anomalous RGB are enhanced in [Fe/H], and have, on average, higher [Na/Fe] values than the standard RGB, in close analogy with what observed in M\,22 and in the other anomalous GCs. As shown in panel d and e of Fig.~\ref{fig:fig2}, Sgr stars have solar sodium abundance relative to iron and define a tail at high metallicity in the histogram of the [Fe/H] distribution.

\section{Comparison with anomalous GCs}
\label{sec:confronto}
 \begin{figure}[b]
 \begin{center}
  \includegraphics[width=4.35in]{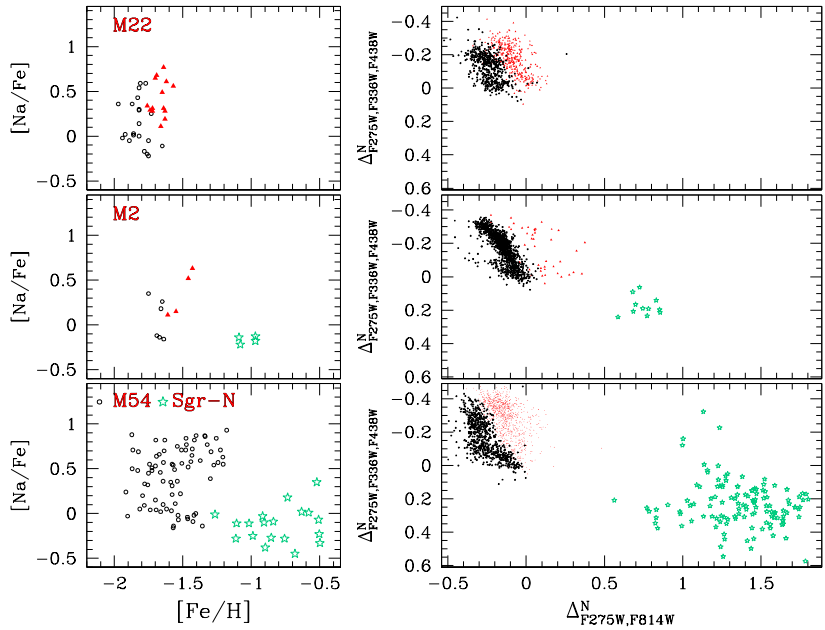}
  \caption{\textit{Left Panels:} [Na/Fe] as a function of [Fe/H] for M\,22 (from Marino et al.\,2009, 2011a, top) M\,2 (from Yong et al.\,2014, middle) and M\,54+Sgr (from Carretta at el.\,2010b, bottom). \textit{Right Panels:} Chromosome map of M\,22, M\,2, and M\,54+Sgr (from Milone et al.\,2015b and in preparation). Stars in the anomalous RGB are colored red while I used aqua starred symbols to represent stars in the Sgr dwarf and in the extreme population of M\,2. }
    \label{fig:fig3}
 \end{center}
 \end{figure}

Further information on multiple stellar population in M\,54 comes from the comparison with anomalous GCs. In Fig.~\ref{fig:fig3} I show [Na/Fe] vs.\,[Fe/H] from high-resolution spectroscopy (left panels) and the chromosome maps derived from the {\it HST} photometry presented by Piotto et al.\,(2015) for two anomalous GCs, namely M\,22 and M\,2 (upper and middle panel) and for the stellar system formed by M\,54 and the Sgr (lower panel). 

M\,22 is the prototype of anomalous GCs and hosts two distinct groups of stars with different metallicity and different abundance of s-process elements and C$+$N$+$O (Marino et al.\,2009; 2011a).
 When compared with M\,22, M\,2 exhibits a more-extreme chemical composition. Indeed it hosts three main stellar components, composed of metal-poor, metal-intermediate, and metal-rich stars (Yong et al.\,2014; Milone et al.\,2015b). While the metal-poor, metal-intermediate population resemble M\,22, the metal-rich population is present in M\,2 only.

The stellar system including M\,54 and the Sgr shares strong similarities with M\,2. M\,54 seems to include populations similar to the iron-poor and iron-intermediate population of M\,2, while the chemistry of Sgr stars is similar to the abundance pattern of the extreme population of M\,2. 
 Unlike the case of $\omega$\,Cen, however, where most of the metal-rich stars are strongly enhanced in [Na/Fe] (Marino et al.\,2011b), both the Sgr stars and the population of M\,2 exhibit low [Na/Fe]. In addition, the extreme population of M\,2 has lower $\alpha$-element abundance than the bulk of M\,2 stars in close analogy with what observed in the Sgr nucleus, where Sgr stars have lower [$\alpha$/Fe] than M\,54. Similar conclusions come from the comparison of the chromosome maps plotted in the right panels of Fig.~\ref{fig:fig3}.

In summary, I have provided evidence that the chromosome maps of GCs are efficient tools to identify anomalous GCs with heavy-element variations. Both spectroscopy and photometry show that the Sgr nuclear cluster, M\,54 shares similarities with anomalous GCs like M\,2, M\,22, NGC\,5286, NGC\,1851, and $\omega$\,Cen. 
 These findings make it tempting to speculate that, similarly to M\,54, the other anomalous GCs are the remnants of dwarf galaxies tidally disrupted by the interaction with the Milky Way.

\begin{acknowledgments}
I thank G.\,Piotto, L.\,Bedin, J.\,Anderson, I.\,King, A.\,Marino, D.\,Nardiello and the other collaborators involved in `The {\it HST} UV Legacy Survey of Galactic Globular Clusters' and A.\,Bragaglia who has revised this manuscript. I acknowledge support by the Australian Research Council through Discovery Early Career Researcher Award DE150101816.
\end{acknowledgments}

\end{document}